\documentclass[twoside,12pt]{article}

\usepackage{times}
\usepackage{amsmath}
\usepackage{alltt}
\usepackage{epsfig}

\newtheorem{lemma}{Lemma}

\newlength{\eztab}
\begin{document}

\pagestyle{myheadings} 
\markboth{AADEBUG 2000}{Slicing of Constraint Logic Programs} 

\title{Slicing of Constraint Logic Programs }

\footnote{In M. Ducass\'e (ed), proceedings of the Fourth International
 Workshop on Automated Debugging (AADEBUG 2000), August 2000, Munich.
  COmputer Research Repository (http://www.acm.org/corr/), 
  cs.SE/0012014; whole proceedings: cs.SE/0010035.}

\author{Gy\"ongyi Szil\'agyi$^1$, Tibor Gyim\'othy$^1$ and Jan 
 Ma{\l}uszy{\'n}ski $^2$  \\
$^1$ Research Group on Artificial Intelligence  Hungarian Academy of Sciences \\
  $^2$ Dept.  of Computer and Information Sci.,
  Link{\"o}ping University, Sweden\\
E-mail:\{szilagyi,gyimi\}@inf.u-szeged.hu, janma@ida.liu.se}


\date{}
\maketitle

\begin{abstract}  
{\bf Abstract. } {\em 
Slicing is a program analysis technique originally developed
for  imperative languages. It facilitates   understanding of data flow
and debugging.

This paper discusses  slicing of Constraint Logic Programs.
Constraint Logic Programming (CLP) is an emerging software technology
with a growing number of applications.
Data flow in constraint  programs is not explicit, and for this reason 
the concepts of slice and the slicing
techniques of imperative languages are not directly applicable.

This paper formulates  declarative notions  of  slice suitable
for CLP. They provide  a basis for defining 
slicing techniques (both dynamic and static) based on variable
sharing. The techniques are further extended by using
groundness information. 

A prototype dynamic slicer of CLP programs implementing
the presented ideas  is briefly described  together with the results of
some slicing experiments. }

\end{abstract}


\input epsf

\newtheorem{theo}{Theorem}

\newtheorem{prop}[theo]{Proposition}

\newtheorem{con}[theo]{Consequence}

\newtheorem{defi}{Definition}

\newtheorem{ex}{Example}

\newcommand{\bd}{\begin{defi}}

\newcommand{\ed}{\end{defi}}

\newcommand{\bt}{\begin{theo}}

\newcommand{\et}{\end{theo}}

\newcommand{\bl}{\begin{lemma}}

\newcommand{\el}{\end{lemma}}

\newcommand{\be}{\begin{ex}}

\newcommand{\ee}{\end{ex}}


\section{Introduction}


This paper discusses slicing of Constraint Logic Programs.
Constraint Logic Programming (CLP) (see e.g. \cite{Mar98})
 is an emerging software technology
with growing number of applications.
Data flow in constraint  programs is not explicit, and for this reason 
the concept of a  slice and the slicing
techniques of imperative languages are not directly applicable.
Also, implicit data flow  makes the 
understanding of program behaviour rather difficult. Thus  
program analysis tools explaining data flow to the user
 could be of great practical importance. This paper presents
 a prototype tool based on the slice concept applied to  CLP.


Intuitively a program {\em slice}
with respect to a specific variable at some program point 
contains all those parts of the program that may affect the value of
the variable ({\em backward slice}) or may be affected by the value 
of the variable ({\em forward slice}). Slicing algorithms
can be classified according to whether they only use statically
available information ({\em static slicing}), or compute those
statements which influence the value of a variable occurrence for
a specific program input ({\em dynamic slice}).
 The slice provides a focus for analysis of the
origin  of the  computed values of the variable in question.
In the context of CLP the intuition remains the same, but
the concept of slice requires precise definition since the nature
of CLP computations is different from the nature of imperative 
computing.

Slicing techniques for logic programs have been discussed in
\cite{Gyi, Duc, Zhao3}.  CLP extends logic programming with 
constraints. This is a substantial extension and the slicing
of CLP program has, to our knowledge, not yet been addressed 
by other authors.
Novel   contributions presented in  this paper are:
\begin{itemize}
\item  {\em A precise formulation of the slicing problem for CLP programs.} 
We first define  a  concept of slice  for a set of constraints,
which is then used to define slices of {\em derivation trees}, representing 
 states of CLP computations.
Then we define  slices  of a program 
in terms of the  slices of its  derivation trees.
\item {\em Slicing techniques for CLP.} 
We present  slicing techniques that make it possible to construct
slices according to the definitions. The techniques are 
based on a simple analysis of variable sharing  and   groundness. 
\item {\em A prototype dynamic slicer.} 
A tool implementing the proposed techniques and some  experiments with its use are briefly described.
\end{itemize}

The  precisely defined concepts of slice gives a solid foundation
for development of slicing techniques. The prototype tool, including
some visualisation facilities, helps the user in better understanding of
 the program and in (manual) search for errors. 
Integration of this
tool with more advanced debuggers is a topic of future work.

The paper is organized as follows.
Section 2 outlines some basic concepts which are then used in
Section 3 to formulate the problem of slicing. 
Section 4 presents and justifies a declarative formalization of CLP slicing,  
based on a notion of dependency relation. 
Section 5 discusses a dynamic backward slicing technique, and the use 
of directionality information
for reducing the size of slices. 
Our prototype tool is described in Section 6, together with
results of some experiments.
Section 7 then  discusses relations to other work. Finally 
in Section 8 we present our conclusions and suggestions for future work.


\section{Constraint Logic Programs}




The  cornerstone of
{\bf Constraint Logic Programming (CLP)} \cite{JaM, Mar98}
is the notion of constraint. Constraints are formulae  constructed
with some {\bf constraint predicates} with a predefined interpretation.
A typical example of a 
constraint is a linear arithmetic  equation
or inequality
with rational coefficients where the
constraint predicate used is equality  interpreted over 
rational numbers, e.g. $X-Y=1$. The variables of a constraint range
over the domain of interpretation.
A {\bf valuation} of a set  $S$ of variables is a mapping $\theta$
from $S$ to the interpretation domain.
A set of constraints  $C$ {\bf is satisfiable } if 
there exists a valuation $\theta$ for the set of variables occurring 
in $C$, such that $\theta (C)$ holds in the constraint domain.   \\

A {\bf constraint logic program} is a set of {\bf clauses}
of the form $h :- b_1,...,b_n$, $n\geq 0$ , where $h,b_1,...,b_n$
are  {\bf atomic formulae}. The predicates used to construct 
$b_1,...,b_n$  are either constraint predicates or other predicates 
(sometimes called {\bf defined  predicates}). The predicate of $h$ 
is a defined predicate. A {\bf goal} is a clause without $h$.
This syntax   extends 
 logic programs  with the possibility of including constraints
into the clauses.
\be
In the  following constraint  program 
the equality constraint
and symbols of arithmetic operations are 
interpreted over the domain of rational numbers.
This simple example is chosen   to simplify 
the forthcoming illustration
of slicing concepts and techniques. The constraints are distinguished
by the curly brackets $\{\}$.

\begin{verbatim}
p(X,Y,Z):- {X-Y=1}, q(X,Y), r(Z).
q(U,V):- {U+V=3}.
r(42).
\end{verbatim}
\ee

Slicing refers to computations. Abstractly, a computation
can be seen as construction of a  tree, from 
renamed instances of clauses. We explain briefly
the  idea discussed formally in   \cite{DM93}. 

Intuitively, a clause $c$ can be seen as a tree with root $h$ {\bf  head}
and leaves $b_1,...,b_n$ {\bf body atoms}. If the predicate of  
$b_i$  appears in the head of a clause $c'$ then a renamed copy $c''$ of
$c'$  can be composed  with $c$ by attaching the head of $c''$ 
to $b_i$. This implicitly adds equality constraints for the corresponding
arguments of the atoms. 
 This process can be repeated for the leaves of 
the resulting tree. More formally this is captured by the following
notion. A {\bf skeleton}  for a program $P$ is a labeled ordered tree with
the root labeled by a goal clause and with the 
nodes labeled by (renamed) clauses of the program; some leaves may instead  be
labeled "?" in which case they are called {\bf incomplete
nodes}.
 Each non-leaf
node has as many children as the non-constraint atoms of its body.
The head predicate of the $i$-th child of a node is the same as the
predicate of the $i$-th non-constraint body atom of the clause labeling
the node.

\begin{figure}[h]
\begin{center}
\epsfbox{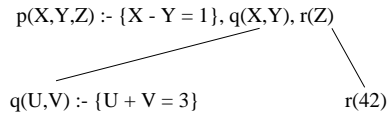}
\caption{A skeleton for the CLP program of Example 1.}
\label{1kep}
\end{center}
\end{figure}

For a given skeleton $S$ the set $C(S)$ of constraints, 
which will be called {\bf the set of constraints of $S$}, consists of :
\begin{itemize}
\item the constraints of all clauses labeling the nodes of $S$
\item all equations $\vec{x} = \vec{y}$ where $\vec{x}$ are the 
arguments of the $i$-th body atom of the
clause labeling a node $n$ of $S$, and $\vec{y}$ are the arguments of 
the head atom of the clause labeling the $i$-th child of $n$. (No equation is created if the
$i$-th child of $n$ is an incomplete node).
\end{itemize}
A {\bf derivation  tree} for a program $P$
is a  skeleton for $P$ whose set of 
constraints is satisfiable. 
If the skeleton is complete (i.e. it has no incomplete node) 
the derivation tree is called a {\bf proof tree}. 
Figure \ref{1kep}  shows a complete  skeleton  tree for the program in
 Example 1.

The  set of constraints of this skeleton is: 
$C(S)= \{X-Y=1, X=U, Y=V, U+V=3, Z=42\}$
and it is satisfiable. Thus the skeleton is a proof tree.

For the presentation of the slicing techniques we need to refer  
to {\bf  program positions} 
 and to {\bf derivation tree positions}. A slice is 
defined  with respect to some particular occurrence of a variable
(in a program or derivation tree), and positions are used to identify
these ocurrences. Positions also identify arguments of the atomic formulae
and their subterms. 
 
    To define the notion of position we assume that some standard
way of enumeration of the nodes of any given tree $T$ is adopted.
The indices of the nodes of $T$ are called  {\bf positions} of $T$
and the set of all positions is denoted $Pos(T)$. Each position
determines a unique subtree of $T$. On the other hand,
 $T$ may have several identical subtrees $T_0$ at different
 positions.

This notation extends also for atomic formulae and terms,
where the positions determine unique subterms.
A position of a term such that the corresponding subterm 
is a variable will be called a {\em variable position}.
We extend the adopted way of enumeration to clauses and 
programs; a {\em program position} is an index in this enumeration
that identifies an atomic formula or a term in a clause of the program.

A single clause may be treated  as a one-clause program.
As discussed above, a derivation tree has its nodes labeled 
by renamed variants of program clauses. By a {\em derivation
tree position} of a derivation tree $T$  
we mean a pair $(i_1,i_2)$, where $i_1$ is a position of the skeleton of $T$ 
and $i_2$ is a position of the clause labeling node $i_1$ in $T$.
The set of all tree positions of a derivation tree $T$
will be denoted by $Pos(T)$. 

Recall  that each label of a derivation tree $T$  is a variant of
a program clause, or of a goal. Therefore the 
positions of $T$ can be mapped in a natural way into the corresponding
program positions.

Similary,  each  occurrence of a variable $X$ 
 in   $C(T)$ (the constraint set of $T$) originates from
a variable position of $X$ in $T$. 
Thus variable positions of $T$ can be linked to the
related constraints of $C(T)$.

Let $\cal P$ be a set of positions of $T$, and  
$\Psi_T({\cal P})$  the set of all variables 
that appear in the terms on positions in  $\cal P$. In this way $\cal P$ 
identifies 
the  subset $C_{\cal P}$ of  $C(T)$ consisting of all 
constraints including variables in $\Psi_T(\cal P)$. 
\be
  Consider the derivation tree of Figure 1.\\  
Let $\cal(P)=\{$ derivation tree positions of the atom $q(X,Y)\} \subseteq Pos(T).$ \\
Then $\Psi_T {\cal(P)} = \{X,Y\}$ and $C_{\cal P}=\{X-Y=1, X=U, Y=V\}. $
\ee


\section{The Slicing Problem}

\begin{figure}[h]
\begin{center}
\epsfbox{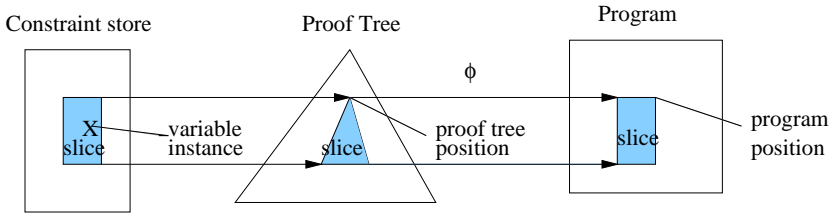}
\caption{A slice of a constraint set, a proof tree and a program.}
\label{slice1}
\end{center}
\end{figure}

Given a variable $X$ in a CLP program we would like to find a fragment 
of the program that may affect the value of $X$. 
This is rather imprecise, hence our objective is to formalize 
this intuition. 
We first define the notion of a slice of  a satisfiable set 
of constraints. A variable $X$ in a derivation tree $T$
has its valuations \cite{JaM, Mar98} restricted by 
the set of constraints of  $T$ (which is satisfiable), 
so our second task will be to define a slice of a derivation tree, 
and finally a slice of a program (see Figure \ref{slice1}).
 
Let $C$ be a set of constraints. Intuitively we would like to remove all
 constraints of the set that do not restrict the valuation of a given variable.
 The binding of a variable 
 $X$  to a value $v$ is said to be a {\em  solution} of $C$ with respect to
$X$ iff there exists a valuation $\nu$ such that $\nu(X)=v$
and $\nu$ satisfies $C$. 
The set of all  solutions of $C$ with respect to $X$ will be denoted by
$Sol(X,C)$.

\bd 
A {\bf slice of a  satisfiable constraint set $C$}
 with respect to $X$  is a subset $S \subseteq C$ such
that $Sol(X,S) = Sol(X,C)$. 
\ed
In other words the set of all solutions  of a slice  $S$ of $C$ 
with respect to $X$ is equal to the set of 
all solutions of $C$ with respect to $X$.
Definition 1 gives implicitly a notion of {\bf minimal slice}:
it is a slice $S$ of $C$ such that if we further reduce $S$ to $S'$,
then $Sol(X,S')$ is different from $Sol(X,C)$. 
Notice that  the whole set $C$ is a slice of itself,
and that the definition does not provide
any hint about how to find a minimal slice. The problem of finding minimal
slices may be undecidable in general, since satisfiability may be undecidable.
So  reasoning about minimality of the constructed slices
seems only be possible in very restricted cases, and for some
specific constraint domains. Our general technique is domain
independent but we show in Section 5 how the groundness
information ( which may be provided by a specific
constraint solver) can be used to reduce the size of the
slices constructed by the general technique.

We now formulate the  slicing problem for derivation  trees.
A derivation tree  $T$ is a skeleton with a set of constraints $C(T)$.
The variables of $C(T)$ originate from positions of $T$. 
Let  $\cal P$ be a set of positions of $T$, i.e. ${\cal P} \subseteq 
Pos(T)$.
Then $\Psi({\cal P})$ identifies the variables of $C(T)$ with
occurrences originating from positions in $\cal P$.
We denote by $C_{\cal P}$  the set of all constraints of  $C(T)$ that include
these variables. 

\bd 
A {\bf slice of a derivation tree} $T$ with respect to 
a variable position of $X$ 
is any subset $\cal P$ of the positions of $T$ 
such that $C_{\cal P}$ is a slice of $C(T)$ with respect to $X$.
\ed

The intuition reflected by this definition is
that the constraints connected with the
positions of the tree not included in a slice  
do not influence restrictions on the
valuation of $X$ imposed by the tree.
We formalize this by referring to the formal 
notion of the slice of a set of constraints.
Notice that any superset of a slice 
is also a slice. 

Finally we define the notion of a CLP program slice
with respect to a variable position. We notice that every position
of a derivation tree $T$ is a (renamed)  copy of a program position
or of  a goal position. This provides a natural map
$\Phi_T$ of the  positions of $T$ 
into program positions  and goal positions. Corresponding to this definition of
$\Phi_T$, for a program position $q$ the set $\Phi^{-1}_T(q)$ contains those proof
tree positions such that if $r \in \Phi^{-1}_T(q)$ then $\Phi_T(r)=q$.

\bd  {\bf A slice of a CLP  program} $P$  with respect to a program position
$q$ is any  set $S$ of positions of $P$ such that 
for every derivation tree $T$ whenever its position $r$
is in $\Phi_T^{-1}(q),$ 
 there exists a  slice $Q$ of $T$ with respect to $r$  such that
 $\Phi_T(Q)\subseteq S$. 
\ed 

This means  that for any derivation tree position $r$, such that $\Phi_T(r)=q$,
and program slice $S$ with respect to $q$, the value of the variable in $r$  
can only be influenced  by variants of the program positions in $S$.



\section{Dependency-based slicing}

The formal definitions of the previous section
  make it possible to state
precisely our objective, which is automatic construction of slices.

Our formulation defines slicing of a CLP program in terms of
the slicing of sets of constraints. 
Generally it is  undecidable whether a subset of
a set of constraints is a slice.
This section presents a rather straightforward sufficient
condition for this.
We provide here a ``syntactic'' approach to  slicing  
 constraint stores, proof trees and programs. 
The propositions follow easily from the definitions
and will be stated without proof. 
More details can be found  in the technical report  \cite{Mal98}.   


\subsection{Slicing  sets of constraints}

We use variable sharing between constraints 
as a basis for slicing  sets of constraints.
Let $C$ be a set of constraints, and $vars(C)$ 
the set of all variables occurring in the  constraints 
in $C$. Let  $X,Y$ be variables in $vars(C)$. 
$X$ is said to {\bf depend explicitly} 
on  $Y$  iff both occur in a constraint $c$ in $C$.
Notice that the explicit dependency relation is symmetric
and reflexive but need not be transitive.

\bd 
A {\bf dependency relation} on $vars(C)$ is
the transitive closure of the explicit dependency
relation.
\ed
The dependency relation on $C$ will be denoted by $dep_C$.
Notice that $dep_C$ is an equivalence relation on $vars(C)$.
We map any   equivalence class $[X]_{dep_C}$ to the
subset $C_X$ of $C$ that consists of all constraints that include variables in $[X]_{dep}$. Then {\bf $C_X$ is a slice of $C$} (see \cite{Mal98}).

\be 
For the set of constraints of 
Example 1 the dependency relation has 
two equivalence classes:
$\{X,Y,U,V\}$ and $\{Z\}$ and gives the
following slice of $C$ with respect to $X$:
\[C_X = \{X-Y=1, \ X=U, \ Y=V, \ U+V=3 \} \]
\ee

\subsection{Slicing of derivation  trees}

We defined the concept of slice for a derivation
tree by referring to the notion of slice of
a set of constraints. To construct slices of
derivation trees we introduce a dependency relation
on the positions of a derivation tree.
We now define a direct dependency relation
$\sim_T$ on $Pos(T)$.  It can be related to the dependency relation on
$vars(C(T))$ \cite{Mal98} and hence can be used for slicing  $T$.

\bd 
Let $T$ be a derivation tree. The direct dependency 
relation $\sim_T$ on $Pos(T)$ is defined as follows:\\
$\alpha \sim _T \beta$  iff  one of the following conditions holds:
\begin{enumerate}
\item $\alpha$ and $\beta$ are  positions in an occurrence
of a clause constraint (constraint edge).
\item $\alpha$ and $\beta$ are  positions in a node equation (transition edge).
\item $\alpha$ and $\beta$ are  positions in an occurrence
of a term (functor edge).
\item $\alpha$ and $\beta$ share a  variable (local edge).
\end{enumerate}  
\ed
Observe that the relation is both reflexive and symmetric.
The transitive  closure $\sim^*_T$ of the direct 
dependency relation will be called the {\em dependency relation} 
on $Pos(T)$. So  $\sim^*_T$ is an equivalence relation.

\begin{prop}
Let $T$ be a proof tree and let $\alpha$ be a variable
position of $T$. Then $[\alpha]_{\sim^*_T}$ is a slice
of $T$ with respect to $\alpha$.
\end{prop}


\subsection{Slicing of CLP  programs}

This section defines a dependency relation on the positions of the program
and then makes use of it in constructing program slices. 
Recall that each  position of a derivation tree $T$ ``originates''
from a position of the selected program $P$. This is formally 
captured by the map $\Phi: Pos(T) \rightarrow Pos(P)$.
The dependency relation $\sim_P$  on $Pos(P)$ we are going to define 
should reflect dependency relations in all proof trees of $P$.
More precisely, whenever $\alpha \sim_T \beta$  
in some tree $T$ we would also like to have $\Phi(\alpha) \sim_P 
\Phi(\beta)$.

\bd 
Let $P$ be a CLP program. The direct dependency relation $\sim_P$ 
on $Pos(P)$ is defined as follows: $\alpha \sim_P \beta$ 
iff at least one of the following conditions holds:

\begin{enumerate}
\item  $\alpha$ and $\beta$ are positions of the same constraint (constraint edge).
\item $\alpha$ is a position of the head atom of a clause
$c$ and  $\beta$ is a position of a body atom of a clause 
$d$  and both atoms have the same 
predicate symbol (transition edge).
\item $\alpha$ and $\beta$ belong to the same argument of a 
function (functor edge).
\item $\alpha$ and $\beta$ are in the same clause and have
a common variable (local edge).
\end{enumerate}
\ed 

The dependency relation of a program can be represented
as a graph.

\begin{figure}[h]
\begin{center}
\epsfbox{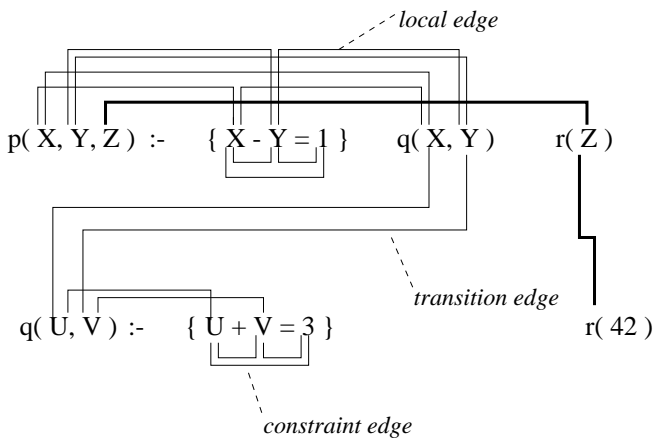}
\caption{Program dependence represented in graphical form  and the backward slice with respect to Z in p(X,Y,Z))}
\label{slice2}
\end{center}
\end{figure}

Comparing the definitions of $\sim_T$ and $\sim_P$ one can
check that   whenever 
$\alpha \sim_T \beta$  
in some tree $T$ of $P$ then  
  $\Phi(\alpha) \sim_P \Phi(\beta)$ as well.
Consequently, for any proof tree $T$ 
 $(\alpha \sim_T^* \beta)  \Rightarrow \Phi(\alpha) \sim_P^* \Phi(\beta)$.
The transitive closure $\sim_P^*$ is an equivalence relation on 
$Pos(P)$.
The following result shows how $\sim_P^*$ can be used for the slicing
of $P$.

\begin{prop}
Let $P$ be a CLP program and let $\beta$ be a position of $P$.
Then $[\beta]_{\sim_P^*}$ is a slice of $P$ with respect to $\beta$.
\end{prop}

Figure \ref{slice2} shows the program dependences and a backward slice of the program  in Example 1  with respect to $Z$ in $p(X,Y,Z)$. For the program in Example 1 $\sim_P^*$ has two equivalence classes.
One of them includes  all occurrences of $Z$ and the 
occurrence of the constant 42. The other consists of  the 
remaining positions.

Definition 6 with Proposition 2 give a method for constructing
program slices without referring to proof trees. 
Thus, we obtain a {\bf static slicing}
technique for CLP. The mapping $\Phi$ was  introduced  to argue about
correctness of this technique. 
 These results also confirm the correctness of the
  slicing algorithm in \cite{Gyi} since a logic program can be viewed  as
   a constraint logic  program. \\
The proposition shows that the concept of dependency relation
on program positions provides a sufficient condition for slicing
a CLP program. However, the slice obtained may be quite large, sometimes 
it may even include the whole constraint store. 
 
We propose two ways for  addressing  this problem. 
On one hand, improvement is possible if we handle the so called 
``calling context problem''\cite{Reps94} which appears when 
the same predicate is called from two different clauses. 
As we explained in \cite{Mal98}  it is possible to adapt to CLP 
the solution proposed by Horwitz et. al \cite{Reps94} for procedural languages. Another way  of reducing the size
 of  slices is to infer and to take into account 
information about proliferation of ground instantiations
during the execution of the program. A  directional
dynamic slicing technique based on this idea 
is presented in the next section.


\section{Dynamic directional  slicing}

All the dependency relations discussed so far were  symmetric.
Intuitively, for a constraint $c(X,Y)$
 a restriction imposed on valuations of $X$ usually
influences admissible valuations of $Y$ and vice versa.
However if $c(X,Y)$ belongs to a satisfiable constraint set $C$
and some other constraints of $C$ make $X$ ground then
the slice of $C$ with respect to $X$ need not include $c(X,Y)$.
For example, if $C = \{X+1=0, Y>X \}$ is interpreted on the integer
domain  then $\{X+1=0\}$ is a slice of $C$ with respect to $X$.
This slice can be constructed by using information about
groundness of variables occurring  in the dependency graph.

This section shows how to use groundness information
in derivation tree  slicing, that is in dynamic slicing
of CLP programs. It extends to CLP the ideas of dynamic
slicing for logic programs presented in \cite{Gyi}.
In our approach groundness
is captured by adding directionality information to dependency graphs.
The directed graphs show the propagation of ground data
during the  execution of  CLP programs, and these graphs can then be used to
 produce more precise slices.
The groundness information will be collected during
the computation that constructs the derivation tree to be
sliced. The proposed concepts are also applicable to the case
of static slicing, where the groundness information has to
be inferred by static analysis of the program. This is however
not discussed in  this paper.

\subsection{Groundness Annotations}

Groundness information associated with a derivation tree will be
expressed as an annotation of its positions.
The annotation classifies the positions of a derivation tree or
the positions of the CLP program.
The positions are classified as
{\em inherited} (marked  with $\downarrow$),
{\em synthesized} ($\uparrow$) and {\em dual} ($\updownarrow$).
An annotation is {\em partial} if some positions are dual. Formally speaking,
 an annotation is a mapping
$\mu$ from the positions into the set
$\{\downarrow,\uparrow,\updownarrow\}$ \cite{DM93}.

The intended meaning of the annotation is as follows.
An  inherited position is a position which is   ground at time of
calling, that is when the equation involving this position
is first created during the construction of the derivation tree.
A synthesized positions is a position  which is  ground at
success, that is when the subtree having the position in its
root label is completed in the computation process.
The dual positions of a proof tree are those for which
no groundness information is given, including  those
which are ground neither at call nor at success.

The annotations will be collected during the execution
of the program. Alternatively, they may be inferred
with some straightforward inference rules discussed in \cite{Mal98}.

We now introduce the following auxiliary terminology relevant to the
annotated positions of a CLP program.
The inherited positions of the head atoms and the synthesized positions of the
body atoms are called {\em input positions}. Similarly, the synthesized
positions of the head atoms and  inherited positions of the body atoms are
called {\em output positions}.
Note that dual positions
are not strictly classified as input or output ones.
Alternatively, if we say that a position is annotated as an
output  we  mean that it is annotated
as inherited provided it is a position in a body atom,
or annotated as synthesized if it is a position of the head of a
clause.

\be Consider the following CLP program:
\begin{verbatim}
1. p(X,Y) :- r(X), q(X,Y).
2. r(3).
3. q(U,V) :- {U+V = 5}.
\end{verbatim}

\ee
The corresponding annotated proof tree for the goal $p(X,Y)$
is presented in Figure \ref{annot1}, where the actual positions have been
replaced
by I (the input positions) and by O (the output positions):

\begin{figure}[h]
\begin{center}
\epsfbox{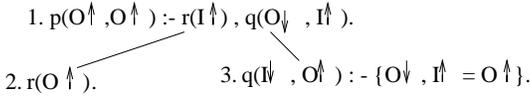}
\caption{The annotated proof tree for Example 5}
\label{annot1}
\end{center}
\end{figure}

The annotation reflects groundness propagation during
the computation, as discussed below.
The variables $X$ and $Y$ of $p(X,Y)$ are annotated as output, since they are
ground at success of $p$, and $p(X,Y)$ is a head atom. For the same reason the
 argument of the fact $r(3)$ in node 2 is annotated as output.
 The variable $U$ in the first argument of the predicate $q(U,V)$ in node 3 is
ground at call, so it is annotated as input, while $V$ is ground at success of
$q$ so it
is annotated as output.


In a CLP program groundness of a position depends generally on
 the used  constraint solver.
Monitoring the execution, also in this case,  we are able to annotate
certain positions as inputs or outputs.
For example the $indomain(X)$ constraint
of CHIP instantiates $X$ to a value in its domain, so that $X$ is an input
position.
Rational solvers can usually solve linear equations. For example in the
rational constraint  $Y=2*X+Z$, $X$ can be annotated input  if $Y$
an $Z$ are output. 


\subsection{Directional slicing  of  derivation trees}

Having input/output information for positions of a derivation tree,
we then add directions to its dependence graph. The following definition
describes how it can be achieved.

\bd {\bf  Directed  Dependency Graph of a Proof Tree} \\
Let  $T$ be a proof tree, $T_G = (TreePos(T), \sim _T)$
its proof tree dependence graph,
then the directed dependence graph of $T(P)$ can be defined as: \\
$T_{DG} = (TreePos(T), \rightarrow_{T})$, where:
\begin{itemize}
\item $\alpha \rightarrow _{T(G)} \beta$ if
$\alpha \sim _{T (G)} \beta $ is a transition edge,
 $\alpha$ is an output  position and $\beta$ is an input
position
\item $\alpha \rightarrow _{T(G)} \beta$
if $\alpha \sim _{T(G)} \beta $ is a local edge,
 $\alpha$ is an input position and $\beta$ is an output position
\item $\alpha \rightarrow _{T(G)} \beta$
and $\beta \rightarrow _{T(G)} \alpha$
in every other case when $\alpha \sim _{T(G)} \beta$
\end{itemize}
\ed

From the definition of $\rightarrow_{T(G)}$
 assuming correctness
of the annotation we find that if $\alpha$ is a  position
of $X$ in $T$ and it is annotated as input or as output then $C(T)$
binds $X$ to a single value.  This value is determined
by the constraints connected with those positions that are in the set
$\{\beta | {\beta} \rightarrow^*_{T(G)} \alpha\}$. Thus we have:

\begin{prop}
$\{\beta | {\beta} \rightarrow ^*_{T(G)} \alpha\}$ is a slice
of $T$ with respect to $\alpha$.
\end{prop}

These slices are usually more precise than in the case when the groundness
information is not used.


The concept of directed dependency graph can be extended
to programs and used for static slicing. This requires good
methods for static groundness analysis to infer annotations
for program positions. Some suggestions for that can be
found in \cite{Mal98}.


\section{A Prototype Implementation  }

\begin{figure}[h]
\begin{center}
\epsfbox{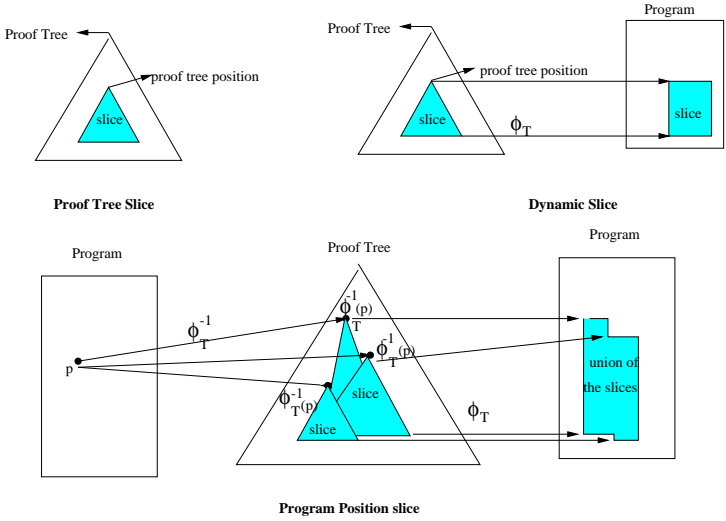}
\caption{Proof Tree, Dynamic and Program Position Slice.}
\label{slice4}
\end{center}
\end{figure}

We  developed a prototype in SICStus Prolog for dynamic backward slicing 
of constraint logic  programs written in SICStus.
The tool  handles a realistic subset of Prolog,
including constructs such as {\em cut, if-then} and {\em  or}.
The {\em inputs} of the slicing system are:  the source code, 
a test case (a goal) and (after the execution) 
the   execution traces given by the Prolog interpreter. 
From this information the 
{\em Directed  Proof Tree  Dependence Graph} (Definition 7)
 may be constructed.  
The following three types of slice algorithms were implemented  
(see Figure \ref{slice4}):
\begin{enumerate}
\item {\bf Proof tree slice} \\
In this case the user chooses  an argument position of the created Proof Tree, 
and the slice is constructed with respect to this proof tree position 
using the Directed Proof Tree Dependency Graph (see Definition 7). 
This kind of slice is useful when the user is interested in the data dependences
  of the Proof Tree.   
\item {\bf Dynamic slice} \\
This case is very similar to 1, but the constructed slice 
of the Proof Tree is mapped back to the program.
This is the 
classic dynamic slice approach \cite{KAR97}, as in the case 
of procedural languages. 
So this slice provides a slice of the program.
\item {\bf Program position slice} \\
In this case the user selects a program position. The system provides all 
instances of this program position in the Proof Tree,  creates 
the proof tree slice for every instance, then the union of these 
slices are constructed and mapped back to the program. So this 
algorithm also provides a slice of the program,  
which  shows all dependences of a program position for a given test case.    
\end{enumerate}
A graphical interface draws the proof tree (see Figure 7), marked with different 
colored  nodes that are in the Proof tree  slice, and 
in the case of a dynamic slice and program position slice the 
corresponding slice of the program is highlighted. The label of the nodes
identify the nodes of the proof tree including the name of the predicate and the
annotation of its arguments.

In the implementation we applied a very simple annotation technique: 
 the inherited positions were those which were ground at the time of calling, 
 while synthesized positions were those which were ground at success.   
This method provides  precise annotation, because we continuously extract information 
 from the actual state of the constraint store. However, in the current
 implementation slicing can only be done on  argument positions. If an argument
  includes several variables it is not possible to distinguish between them, 
  which makes 
 the constructed slice "less precise" (compared to the minimal slice).
 So, our aim is to improve the
  existing implementation, extend it to variable positions.

In the present version of the tool
the sliced proof tree corresponds to the first
success branch of the SLD tree. As the proof tree slice definition
is quite general,  there is no real difficulty in applying the
technique to all success  branches of SLD tree.
The extension to failure  branches is discussed in 
 \cite{Deb99}. Currently we are working on an implementation of these extensions.

Systematic slicing  experiments were performed on a number
of  constraint logic programs (written in SICSTus Prolog).
Each of them was  executed with a number  of test inputs
to collect data about the relative size of a slice with respect 
to the proof tree, depending on the choice of the position.
The selected application programs \cite{JaM, Mar98} had different
 language structures (use of cut, or, if-then, databases, compound 
constraint), and were of different size.

\begin{figure}[h]
\begin{center}
\epsfbox{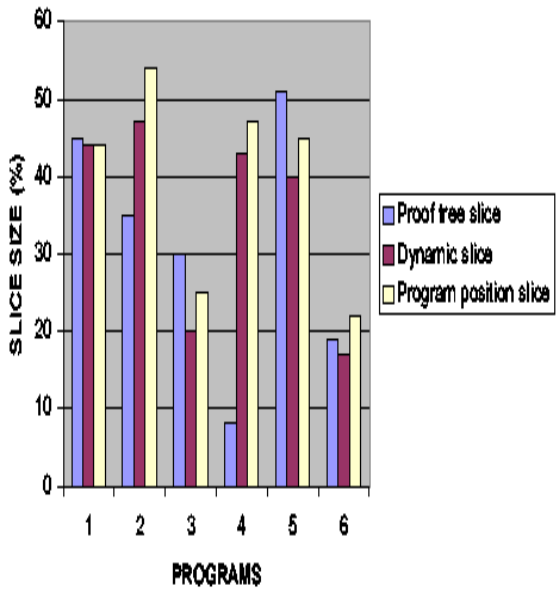}
\caption{Comparison of the Proof Tree Slice,  Dynamic Slice and the Program Position Slice.}
\label{slices}
\end{center}
\end{figure}

The summarized data of the test results on proof tree slicing
is listed in {\em Table 1}. 
The comparison of the three kind of slices with respect to the number of nodes
are shown in Figure \ref{slices}.

\begin{table}[!h]
{\tiny
\begin{tabular}{|ll|c|c|c|lr|lr|} \hline
 &PROGRAM&NUMBER OF&NUMBER OF &NUM. OF COMP-& AVERAGE &SIZE OF&AVERAGE& SIZE OF \\ 
 &&CLAUSES&TEST CASES&UTED SLICES&THE PROOF& TREE&SLICES \\ \hline
 & & & & & NODE & ARG. POS.& NODE  & ARG.POS.\\ \hline
 
 1 & LIGHTMEAL  & 11 & 1 & 17  &  9    & 15     & 44.64 \% & 41.17 \% \\ \hline
 2 & CIRC       & 5  & 2 & 139 & 33.35 & 54.09  & 35.14 \% & 54.44 \% \\ \hline
 3 & SUM        & 6  & 4 & 626 & 80.26 & 120.77 & 30.46 \% & 49.34 \% \\ \hline
 4 & FIB        & 4  & 6 &1349 &399.57 & 533.51 &  7.56 \% &  8.53 \% \\ \hline
 5 & SCHEDULING & 20 & 1 & 575 & 134   & 390    & 51.21 \% & 71.66 \% \\ \hline
 6 & PUZZLE     & 30 & 2 &1363 &227.57 & 607.82 & 19.05 \% & 14.93 \% \\ \hline
\end{tabular}
  }
\caption{Proof Tree Slice}
\end{table}

\begin{figure}[h]
\begin{center}
\epsfbox{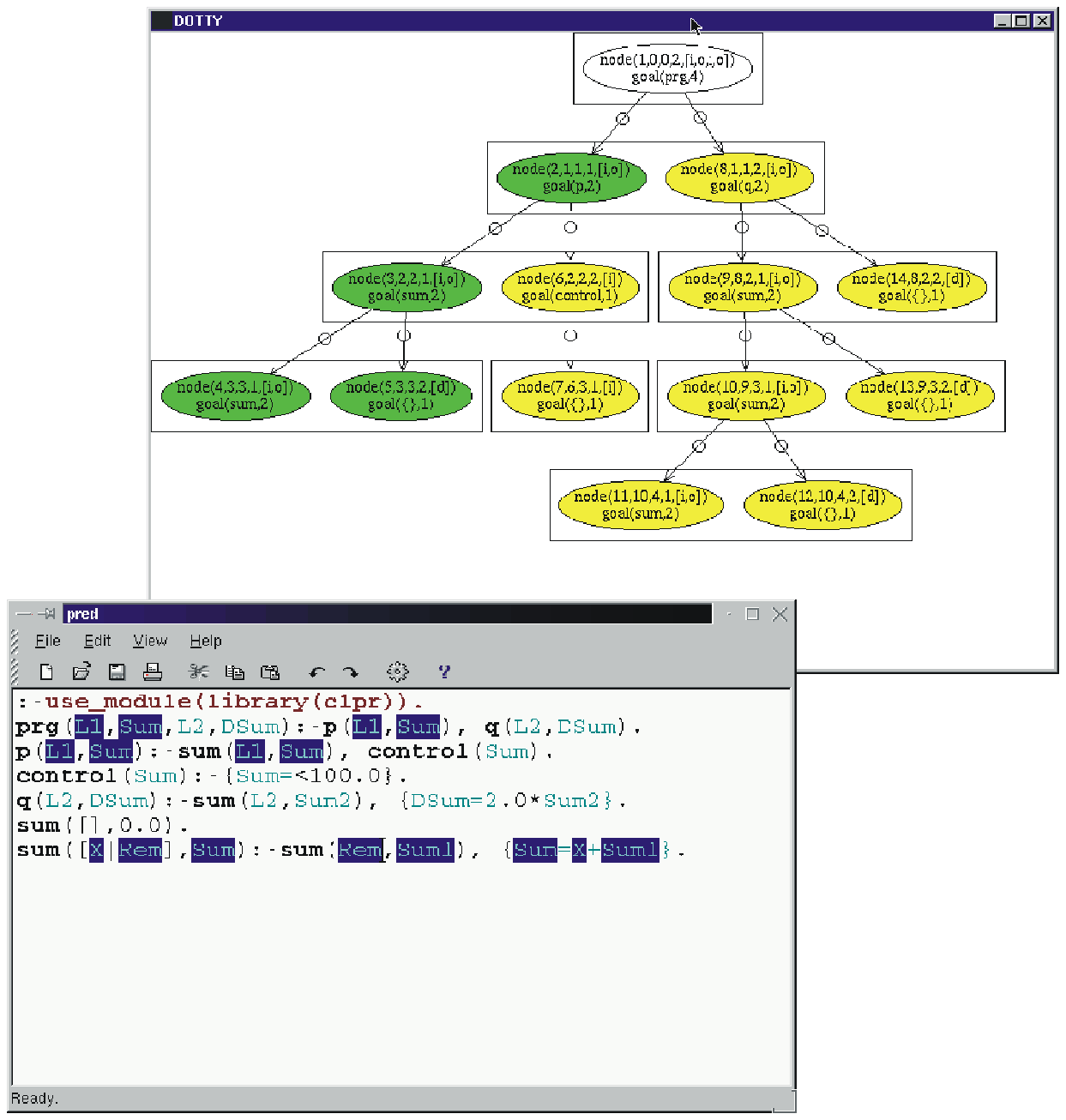 }
\caption{Displayed program and proof tree slices.}
\label{elso1 kep}
\end{center}
\end{figure}

It should be mentioned that 
 the average slice size (in percent) in these experiments 
  had no correlation with the size of the program. 
The average slice size was $32\%$ of 
the number of  executed nodes and $40\%$ of the executed argument positions.

The intended application of our slicing method  is 
to support debugging of constraint programs.
A bug in a program shows up as a symptom during the
execution of the program on some input. 
This means that in some computation step
(i.e. in some derivation tree)
a variable of the program is bound in a way 
that does not conform to user expectations.
A slice of the derivation tree with respect to this occurrence
of the variable can be mapped into the text of the program
and will identify a part  of the program 
where the undesired binding was produced.
This would provide a focus for debugging,
whatever is the debugging technique used.
Intersection of the slices produced for different
runs of the buggy program may further narrow the focus.

In particular, our slicing technique can be combined
with  Shapiro's algorithmic debugging \cite{Shap},
designed originally for logic programs. 
As shown in \cite{IDTS} the slice  of a derivation tree 
of a logic program  may  reduce the number of queries of
algorithmic debugger.  
This technique can also be applied to
the case of algorithmic debugging of CLP programs
discussed in [19].


\section{Related Work}

Program slicing has been widely studied for imperative programs 
\cite{GBF, KAM93, HR92, KAR97, Reps94}. To our knowledge only 
a few papers have dealt with the problem of slicing logic programs,
 and  slicing of
constraint logic programs has not been investigated till now.  

We provided a theoretical basis for the slicing of  proof 
trees and programs starting from a semantic definition of the 
constraint set dependence. This applies as well to the special case
of logic programs (the Herbrand domain). In particular  it  justifies 
the slicing technique of Gyim\'othy and Paakki  \cite{Gyi} 
developed to reduce the number of queries for algorithmic debugger
and  makes it possible to extend  the latter to the general case of CLP.
The directional slicing of Section 5 is an extension of this
technique to the general case of CLP.

 Schoening and Ducass\'e \cite{Duc} proposed a backward slicing 
algorithm for Prolog which produces executable slices. For a target 
Prolog program they defined a slicing criterion which consists of a 
goal of $P$ and a set of argument positions $P_a$, along with a slice 
$S$  as a reduced and executable program derived from $P$. 
An executable slice is usually less precise but it may be used
for additional test runs. Hence the objectives of their work are somewhat
different from ours, and
their algorithm is  applicable to a limited subset of Prolog 
programs.  

 In \cite{Zhao3} Zhao et al defined some static and dynamic 
slices of  concurrent logic programs called  literal dependence nets. 
They presented a new program representation 
called the argument dependence net for concurrent logic programs to 
produce static slices at the argument level. There are some 
similarities between our slicing techniques and Zhao's methods,
since we also  rely on some dependency relations. However, 
the focus of our work has been on CLP not on concurrent logic programs,
and  our main aim has been a declarative formulation of the slicing problem 
whic provides a clear reference basis for 
proving the correctness of the proposed slicing methods.

The results of  the ESPRIT Project DiSCiPl \cite{DHM00} show the importance
of visualisation in the debugging of constraint programs. Our tool
provides a rudimentary visualisation of the sliced proof tree.
It was pointed out by Deransart and Aillaud \cite{DA00}
that abstraction techniques
are needed in the visualisation of the search space.
Program position slicing, if applied
to all branches of the SLD-tree,  provides  
yet another abstraction of the search space.

\section{Conclusions}

The paper offers a precise declarative formulation of the slicing problem
for CLP. It also  gives a solid  reference basis 
for deriving various techniques of slicing  CLP programs
in general and  for logic programs treated as a special case.
This technique was illustrated  by deriving 
 the directional data flow slicing  technique  for CLP,
which is an extension of \cite{Gyi} applied to CLP. As a side effect, the latter, which was presented
in somewhat pragmatic setting,  obtains a theoretical 
justification.

The paper presents also a prototype slicing tool using
this technique. The experiments with the tool show that
the obtained slices were  quite precise in some cases,
and on average provided a substantial reduction of the
program.

The future work  will focus on the application 
 of slicing techniques to
constraint programs debugging. Two aspects being considered. 
Firstly, in the manual debugging of CLP programs it is necessary
to support user with a tool that facilitates the understanding
of the program. We believe that a future version of our slicer
equipped with suitable  visualisation features  could 
be very suitable for this purpose. 
Secondly, the  slicing of a derivation tree
can often reduce the number of queries in 
algorithmic debugging of logic programs \cite{Shap}. 
(The algorithmic debugging technique has been extended
to CLP  \cite{TF00}). 

 As a first step in this direction we  are going to integrate 
our CLP slicing 
tool  with the IDTS algorithmic debugger \cite{IDTS}, originally  
developed for pure logic programs.\vspace{1.5cm} 

{\bf Acknowledgments} \\

The work of the first and second authors was supported by the grants 
 OTKA T52721 and IKTA 8/99.


\bibliographystyle{plain}

\begin{thebibliography}{99}




 



\bibitem{DA00} P. Deransart and C. Aillaud: {\em Towards a Language
for CLP Choice-tree Visualisation}, In:
 P.~Deransart, M.~Hermenegildo, and J.~Ma{\l}uszy\'nski, (editors),
  {\em Analysis and Visualization Tools for Constraint Programming}, LNCS.
  Springer Verlag, 2000 (to appear).

\bibitem{DHM00}
P.~Deransart, M.~Hermenegildo, and J.~Ma{\l}uszy\'nski, (editors),
  {\em Analysis and Visualization Tools for Constraint Programming}, LNCS.
  Springer Verlag, 2000 (to appear).

\bibitem{DM93} P. Deransart and J. Ma{\l}uszy{\'n}ski: 
{\em A grammatical view of logic programming}. The MIT Press 1993.


\bibitem{GBF} T. Gyim\'othy, \'A. Besz\'edes and I. Forg\'acs: An Efficient 
Relevant Slicing Method for Debugging. {\em  In Proceedings of 7th European 
Software Engineering Conference (ESEC'99)}, LNCS 1687 Springer Verlag,
 pages 303-322, Toulouse, France, September 1999. 

\bibitem{Gyi} T. Gyim\'othy and J. Paakki: Static Slicing of Logic 
Programs. {\em In Proceedings of Second International Workshop on Automated and 
Algorithmic Debugging (AADEBUG'95)}, pages 85-105, Saint Malo, France, May 1995.

\bibitem{Deb99} Harmath L., Szil\'agyi Gy., Gyim\'othy, T., Csirik J.: Dynamic Slicing of Logic Programs. {\em Program Analysis and verification, Fenno-Ugric
Symposium (FUSST'99)}, pages 101-113, Tallin, Estonia 1999. 
 
\bibitem{Reps94} S. Horwitz, T. Reps and D. Binkley: Interprocedural Slicing Using Dependence Graphs. {\em In Proceedings of ACM Transactions on Programming Languages and Systems 12}, pages 26-61, 1990.

\bibitem{HR92} S. Horwitz and T. Reps: The Use of Program Dependence 
Graphs in Software Engineering. {\em In Proceedings of the Fourteenth International 
Conference on Software Engineering }, pages 
392-411, Melbourne, Australia, May 1992.
  
\bibitem{JaM} J.Jaffar and M.J.Maher: Constraint logic programming: A 
survey.
{\em The Journal of Logic Programming} 19/20:503-582, 1994.

\bibitem{KAM93} M. Kamkar  and P. Fritzson: Evaluation of Program 
Slicing tools. {\em In 2nd International Workshop on Automated and 
Algorithmic Debugging (AADEBUG'95) }, pages 51-69, Saint Malo, France, 
May 1995.

\bibitem{IDTS} G. K\'okai, L Harmath, T. Gyim\'othy: Algorithmic Debugging and
Testing of Prolog Programs, {\em  In Proceedings of the Fourteenth International
Conference on Logic Programming, Eighth Workshop on Logic Programming
Environments (ICLP'97)}, pages 14-21, Leuven, Belgium, September 1997. 

\bibitem{KAR97} B. Korel and J. Rilling: Application of Dynamic 
Slicing in Program Debugging. {\em  In Proceedings of the Third 
International Workshop on Automatic Debugging (AADEBUG '97)}, 
Link\"oping, pages 43-59, Sweden, May 1997. 

\bibitem{Mar98} K. Marriott and P.J. Stuckey: Programming with Constraints.
An Introduction. The MIT Press, 1998







\bibitem{Duc} S. Schoening and M. Ducass\'e: A Backward Slicing 
Algorithm for Prolog.  {\em In Proceedings of Third International Static Analysis  Symposium SAS'96}, LNCS 1145, 317-331, Springer-Verlag 1996.


\bibitem{Shap} E. Shapiro: Algorithmic Debugging. {\em The MIT 
Press}, 1983.

\bibitem{Mal98} Gy. Szil\'agyi, T. Gyim\'othy, J. Maluszy\'nski:
Slicing of Constraint Logic Programs. Technical Report, {\em Link\"oping University
Electronic Press} 1998/020, {\em www.ep.liu.se/ea/cis/1998/002}.



\bibitem{TF00} A. Tessier and G. F{\` e}rrand. {\em Declarative Diagnosis in 
the CLP Scheme}, In:
 P. Deransart, M. Hermenegildo, and J.~Ma{\l}uszy\'nski, (editors),
  {\em Analysis and Visualization Tools for Constraint Programming}, LNCS.
  Springer Verlag, 2000 (to appear).

\bibitem{Tip} F. Tip: A survey of Program Slicing Techniques. {\em 
Journal of Programming Languauges}, Vol.3., No.3, pages 121-189,
September, 1995.
 


\bibitem{Zhao3} J. Zhao, J. Cheng and K. Ushijima: Slicing Concurrent 
Logic Programs. {\em In Proceedings of Second Fuji International Workshop on 
Functional and Logic  Programming}, pages 143-162, 1997.


\end{thebibliography}

\end{document}